\documentclass[11pt,twoside]{article}

\usepackage{asp2004}
\usepackage{epsf}
\usepackage{psfig}
\usepackage{epsfig}
\usepackage{lscape}

\begin{document}
\title{Spitzer Massive Lensing Cluster Survey}   
\author{
E.~Egami$^{1}$,
G.~H.~Rieke$^{1}$,
J.~R.~Rigby$^{1}$,
C.~Papovich$^{1}$,
J.-P.~Kneib$^{2}$,
G.~P.~Smith$^{3}$,
E.~Le Floc'h$^{1}$,
K.~A.~Misselt$^{1}$,
P.~G.~Per\'{e}z-Gonz\'{a}lez$^{1}$,
J.-S.~Huang$^{4}$,
H.~Dole$^{5}$, \&
D.~T.~Frayer$^{6}$
}   
\affil{(1) Univ.\ of Arizona, (2) Laboratoire d'Astrophysique de
  Marseille, (3) Caltech, (4) CfA , (5) Institut d'Astrophysique
  Spatiale, (6) SSC}

\begin{abstract} 
We are currently undertaking a {\em Spitzer} GTO program to image
$\sim 30$ massive lensing clusters at moderate redshift with both IRAC
and MIPS.  By taking advantage of the gravitatinoal lensing power of
these clusters, we will study the population of faint galaxies that
are below the nominal {\em Spitzer} detection limits.  Here, we
present a few examples of our science programs.
\end{abstract}



\vspace*{-1cm}

\section{Introduction}

The gravitational lensing power of massive galaxy clusters can be
exploited to improve the nominal {\em Spitzer} detection limits.
Clusters at $z=0.2-0.4$ typically magnify a background sky area of
$\sim$ 1~arcmin$^{2}$ with a magnification factor of $\ga 5$ and often
up to 20--30, which directly translates into a gain in the
signal-to-noise ratio.  Furthermore, if the sensitivity is limited by
source confusion, gravitational lensing offers a further benefit of
reducing the confusion noise by spreading out the background galaxies.
For these reasons, Spitzer imaging of massive cluster fields could
potentially achieve a depth that is not attainable even in the GOODS
fields.  In the past, the effectiveness of such a strategy was
anticipated for submillimeter observations \citep{Blain97}, and was
susscessfully demonstrated by a series of SCUBA cluster observations
\citep[e.g.,][]{Smail02}.

Motivated by this potential and the success of the submillimeter
cluster surveys, we are currently undertaking a Spitzer GTO program to
image $\sim 30$ massive clusters with both IRAC and MIPS.  The main
target selection criteria are the following: (1) X-ray luminous, i.e.,
massive ($L_{X} > 7 \times 10^{44}$ erg/s); (2) moderate redshift ($z
\sim 0.15-0.5$); (3) low IR background ($N_{H} < 3.5 \times 10^{22}$
cm$^{-2}$); and (4) abundance of ancillary data (e.g., {\em HST}
images, which are necessary to construct accurate cluster mass
models).  All the clusters will be imaged in the four IRAC bands and
in the MIPS 24 $\mu$m band; MIPS 70 and 160 $\mu$m observations will
be carried out for $\sim 10$ clusters.

Such a rich data set provides opportunities for many exciting
scientific projects.  Here, we will present a few examples,
concentrating on the lensed sources.  In a parallel effort, cluster
galaxy evolution is also being studied in combination with a sample of
local and high-redshift ($z \sim 1$) clusters.

\section{24 $\mu$m Sources in the Cluster Cores}

\begin{figure}[!t]
  \vspace*{-1.7cm}
  
  \hspace*{-1cm}\epsfig{file=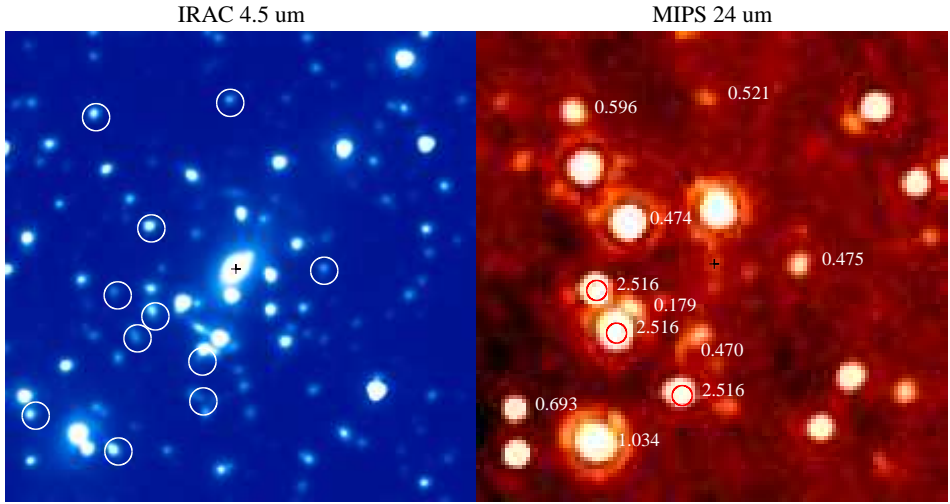,angle=90,width=5.6in}

  \vspace*{-2.3cm}

  \caption{Spitzer images of Abell 2218: Left -- IRAC 4.5 $\mu$m
  image.  Galaxies with spectroscopic redshifts \citep{Ebbels98} are
  marked with circles; Right -- MIPS 24 $\mu$m image.  The
  spectroscopic redshifts are listed.  The triply lensed images of the
  $z \sim 2.5$ submillimeter galaxy \citep{Kneib04a} are marked with
  circles. In both images, the position of the cD galaxy is marked
  with a cross. \label{a2218}}

  \vspace*{-0.4cm}

\end{figure}

Lensing clusters are especially useful for deep imaging at 24 $\mu$m
because the cluster cores, where the lensing magnification is
strongest, are dominated by early-type galaxies, which are faint at 24
$\mu$m.  As shown in Figure~\ref{a2218}, the majority of the 24 $\mu$m
sources seen in the cluster core are background lensed sources.

Scientifically, the driver here is to study the population of faint 24
$\mu$m galaxies which are below the detection limit of field deep
surveys.  Most recently, the evolution of 24 $\mu$m-selected galaxies
have been investigated up to $z=1$ \citep{LeFloch05} and to $z=2.5$
\citep{Perez05} based on the Spitzer GTO deep survey data of CDF-S and
HDF-N using both spectroscopic and photometric redshifts.  However,
these 24 $\mu$m data are not quite deep enough to constrain the shape
of the luminosity function below $L^{*}$ at $z \ga 1$.  Furthermore,
many of the high-redshift 24 $\mu$m galaxies are too faint to detect
at 70 and 160 $\mu$m, making it impossible to determine their total IR
luminosities directly.  This cluster program should help address these
problems by detecting fainter 24 $\mu$m sources and by increasing
70/160 $\mu$m detections.

Figure~\ref{a2218} also shows the detection of the triply lensed $z
\sim 2.5$ submm galaxy discovered by \citet{Kneib04a}.  The three
lensed images are faint in the optical, but they are clearly detected
in all four IRAC bands and are fairly bright at 24 $\mu$m (though
still below 1 mJy).  Curiously, the spectral energy distribution (SED)
measured by IRAC is different from what we saw with the
submillimeter/radio sources in the Lockman Hole \citep{Egami04}.
Considering that this source has a high magnification (a total of
$\times 45$ with all three images) and that each image further breaks
up into substructures with different colors, the IRAC SED might be
distorted by differential magnification.

Although most cD galaxies are faint at 24 $\mu$m as is the case with
Abell 2218, a few cD galaxies in the so-called cooling flow clusters
show 24 $\mu$m luminosities one to two orders of magnitude larger than
those of average cD galaxies.  This indicates that these cD galaxies
harbor a luminosity source which is blocked from our view at shorter
wavelengths.

\clearpage

\section{Properties of Stellar Populations at $z=4-5$ and Beyond}

The Balmer/4000\AA\ breaks get redshifted into the $K$ band at $ z
\sim 4$.  Therefore, to determine the properties of the underlying
stellar population for galaxies at $z \ga 4$ (e.g., age, mass), it is
essential to measure fluxes redward of the $K$-band.  Presently, IRAC
is the only instrument that is sensitive enough to provide such
information.

\begin{figure}[h]

  \vspace*{-2.7cm}

  \hspace*{1cm} \epsfig{file=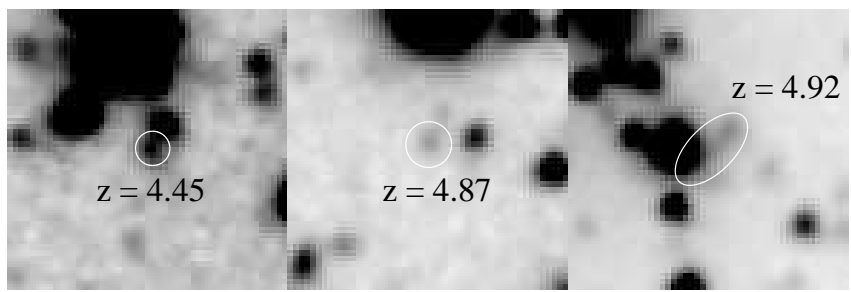,angle=90,width=5in}

  \vspace*{-2.8cm}

  \caption{IRAC 3.6 $\mu$m images of three galaxies at $4<z<5$:
  Left -- $z=4.45$ galaxy in Abell 2219; Middle -- $z=4.87$ galaxy in
  Abell 1689; Right -- $z=4.92$ galaxy in MS1358. The first two
  redshifts are from \citet{Frye02} while the third one is from
  \citet{Franx97}. \label{highz}}

\end{figure}

Figure~\ref{highz} shows the IRAC 3.6 $\mu$m detections of three
galaxies with spectrospic redshifts of $4.5-5$.  The magnification
factors estimated for these galaxies range from 7 to 10, suggesting
that these galaxies would not have been detected without the cluster
lensing.

\begin{figure}[b]
  \vspace*{-3cm}

  \hspace*{-1cm} \epsfig{file=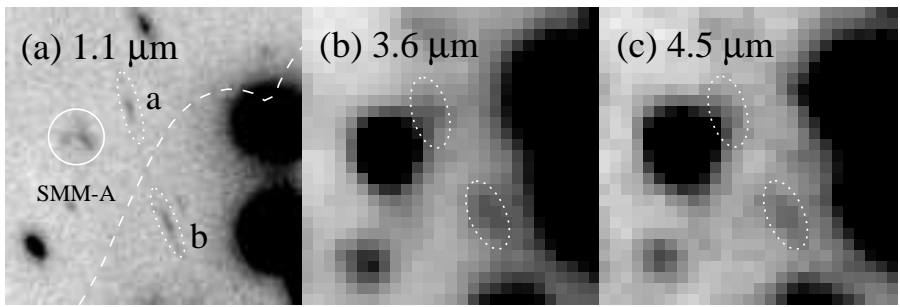,width=5.5in}

  \vspace*{-3cm}

  \caption{{\em HST}/NICMOS and {\em Spitzer}/IRAC images of the
  $z\sim7$ lensed pair: (a) NICMOS 1.1 $\mu$m image.  Components $a$
  and $b$ as well as the $z \sim 2.5$ submillimeter source SMM-A (the
  northernmost of the three lensed images in Figure~\ref{a2218}) are
  marked.  The dashed line indicates the $z \geq 6.5$ critical lines;
  (b) IRAC 3.6 $\mu$m image; (c) IRAC 4.5 $\mu$m image. \label{z7im}}

\end{figure}
The remarkable potential of {\em Spitzer} to probe even higher
redshift was demonstrated dramatically by the IRAC detection of a
gravitationally lensed $z \sim 7$ galaxy at 3.6 and 4.5 $\mu$m
\citep{Egami05}.  This galaxy was first discovered by
\citet{Kneib04b}, and is located in the cluster Abell 2218
(Figure~\ref{a2218}).  Figure~\ref{z7im} shows the {\em HST}/NICMOS
and {\em Spitzer}/IRAC images of the lensed pair. It is located
symmetrically with respect to the critical lines for background
sources at $z \geq 6.5$, suggesting that the redshift must be at least
this high.

Figure~4 compares the SED of component $b$ with a range of
models. Although the redshift has not been confirmed
spectroscopically, photometric redshifts derived from Figure~4 range
from 6.6 to 6.8, consistent with the lower limit set by the lens
model.  The figure clearly illustrates the power of IRAC to measure
the flux longward of the Balmer break for such a high-redshift galaxy.
For this galaxy, we concluded that its underlying stellar population
is in the post-starburst stage with an age of at least $\sim$ 50 Myr
(and quite possibly a few hundred Myr), suggesting that a mature
system is already in place at this early era.  The stellar mass is
$\sim 10^{9}$ M$_{\odot}$, an order of magnitude smaller than typical
Lyman break galaxies at $z=3-4$.

Considering that this galaxy appears to be in the poststarburst stage,
its predecessor is likely to be more luminous at higher redshift,
mitigating perhaps the effect of the increased luminosity distance.
This presents a further exciting possibility that we may be yet to
witness even higher redshift galaxies with {\em Spitzer}.  

\vspace*{0.5cm}

\begin{minipage}{.5\linewidth}

  \hspace*{-1cm} \epsfig{file=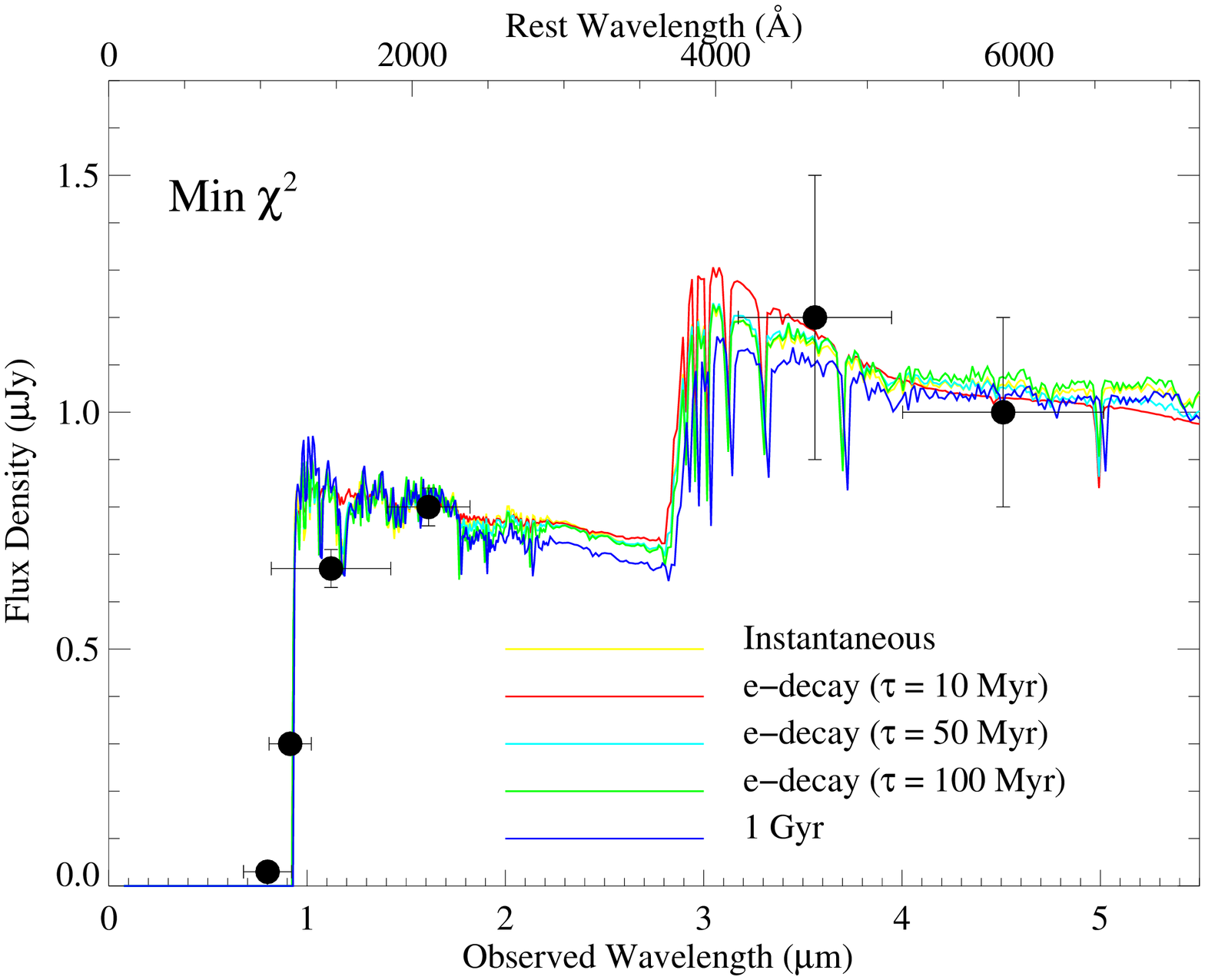,width=2.5in,angle=90}

\end{minipage}\hfill
\begin{minipage}{.4\linewidth}

  {\small Figure~4. SED model fits to the observed SED of the $z \sim
  7$ galaxy (component $b$).  The best (i.e., minimum $\chi^{2}$)
  model for each star formation history is plotted. The restframe
  wavelength at $z = 6.65$ is also shown.  The measured flux at 1.1
  $\mu$m is expected to be significantly lower than the true continuum
  level because the F110W filter passband extends below 1216 \AA\ in
  the restframe.  The intrinsic flux densities (i.e., without the
  magnification) is estimated to be 25 times lower.}

\end{minipage}

%





\end{document}